\documentclass[aps,prl,twocolumn,email,reprint]{revtex4}
\usepackage{amssymb}
\usepackage{amsmath}
\usepackage{natbib}
\usepackage{graphicx}
\usepackage{bbold}
\usepackage{color}
\usepackage{soul}
\usepackage{pdfpages}

\begin{document}
    
    \title{Multimode fiber based single-shot full-field measurement of optical pulses}
    \author{Wen Xiong}
    \author{Shai Gertler}
    \author{Hasan Y{\i}lmaz}
    \author{Hui Cao} 
    \email{hui.cao@yale.edu}
    \affiliation{Department of Applied Physics, Yale University, New Haven, Connecticut 06520, USA}
    \date{\today}
    \begin{abstract}
        
        Multimode fibers are widely explored for optical communication, imaging and sensing applications. The interference of fiber guided modes generates a speckle pattern, which has been used for high-precision spectroscopy, temperature, and strain sensing. Here we demonstrate a single-shot full-field temporal measurement technique based on a multimode fiber. The complex spatiotemporal speckle field is created by a reference pulse propagating through the fiber, and it interferes with a signal pulse. From the time-integrated interference pattern, both the amplitude and the phase of the signal are retrieved in spectral and temporal domains. The simplicity and high sensitivity of our scheme illustrate the potential of multimode fibers as versatile and multi-functional sensors. 
        
    \end{abstract}
    \maketitle
    \section{Introduction}
    A multimode fiber (MMF) can be considered as a complex photonic structure \cite{Rotter17,Mosk12} with diverse degrees of freedom in space, time, spectrum and polarization. As these degrees of freedom are coupled,  an MMF provides a versatile and multi-functional platform for communication \cite{Richardson13,Kahn17}, imaging \cite{Papadopoulos12,Caravaca-Aguirre13,Ploschner15,French18} and sensing applications \cite{Fujiwara12,Wang06,Liu07, Redding12,Redding14,Valley16,Sefler18}. The abundant spatial degrees of freedom have been utilized for controlling linear \cite{Carpenter15,Xiong16,Xiong17,Ambichl17} and nonlinear light propagation \cite{Wright15,Florentin17,Tzang18,Chekhovskoy18,Krupa19} in an MMF. The spatial, temporal, spectral or polarization states of transmitted light are manipulated by shaping the spatial wavefront of an incident beam. Hence, an MMF can function as a microscope \cite{Papadopoulos12,Caravaca-Aguirre13,Ploschner15,French18}, a reconfigurable waveplate \cite{2018_Xiong_LSA} or a pulse shaper \cite{Carpenter15,Xiong16,Xiong17,Ambichl17}. In particular, the coupling between spatial and temporal degrees of freedom in an MMF enables tailoring the output state in time by manipulating the input state in space. However, the reverse process, i.e., extracting the input temporal shape from the output spatial profile of an MMF, has not been explored. It will open the possibility of using an MMF for temporal pulse measurement. 
    
    MMFs have already been employed for various sensing applications. For example, an MMF can be implemented to detect changes in temperature, refractive index and strain \cite{Okamoto88,Fujiwara12,Wang06,Liu07}, because the speckle pattern, produced by the interference of guided modes, is sensitive to external perturbations. For optical coherence tomography (OCT), random temporal speckles generated by MMFs are used to image axial reflectivity profiles \cite{Villiger17}. Furthermore, the dependence of the output spatial pattern on the input spectrum is utilized to transform an MMF into a compact and high-resolution spectrometer \cite{Redding12,Redding14}. However, only the spectral amplitude can be extracted in the spatial intensity pattern, not the spectral phase, which is needed for full-field temporal measurement. 
    
    In this work, we propose and realize a novel method based on an MMF for single-shot full-field measurement of optical pulses. It utilizes the complex yet deterministic spatiotemporal speckle field $E(\mathbf{r},t)$ produced by a reference pulse $f(t)$ propagating through an MMF. Such field $E(\mathbf{r},t)$, which is two-dimensional (2D) in space $\mathbf{r}$ and one-dimensional (1D) in time $t$, interferes with the unknown field $g(t)$ of a signal pulse that is mutually coherent with $f(t)$. The interference pattern is integrated in time by a camera. From this pattern, both the spectral amplitude and the phase of the signal are retrieved. The Fourier transform gives the full field of $g(t)$. The temporal resolution $\delta t$ is set by the temporal speckle size, which is inversely proportional to the spectral bandwidth of the reference pulse $\delta \omega$. The temporal range of a single-shot measurement, $\Delta t$, is set by the temporal length of the transmitted waveform, which is given by the inverse of the spectral correlation width $\Delta \omega$ of the MMF. A fiber with stronger modal dispersion has faster spectral decorrelation, thus covering a longer time window. 
    
    Our scheme can be considered as parallel ghost imaging in time \cite{Devaux16}. Compared to the conventional ghost imaging that relies on the sequential generation of different temporal waveforms \cite{Shirai10,Chen13,Ryczkowski16}, the MMF simultaneously creates many distinct temporal speckle patterns, each at a different spatial location of the output facet, to sample the signal. The parallel sampling enables single-shot measurement, eliminating the requirement for repetitive signals.
    
    \begin{figure*}[t]
        \includegraphics[width=2\columnwidth,keepaspectratio,clip]{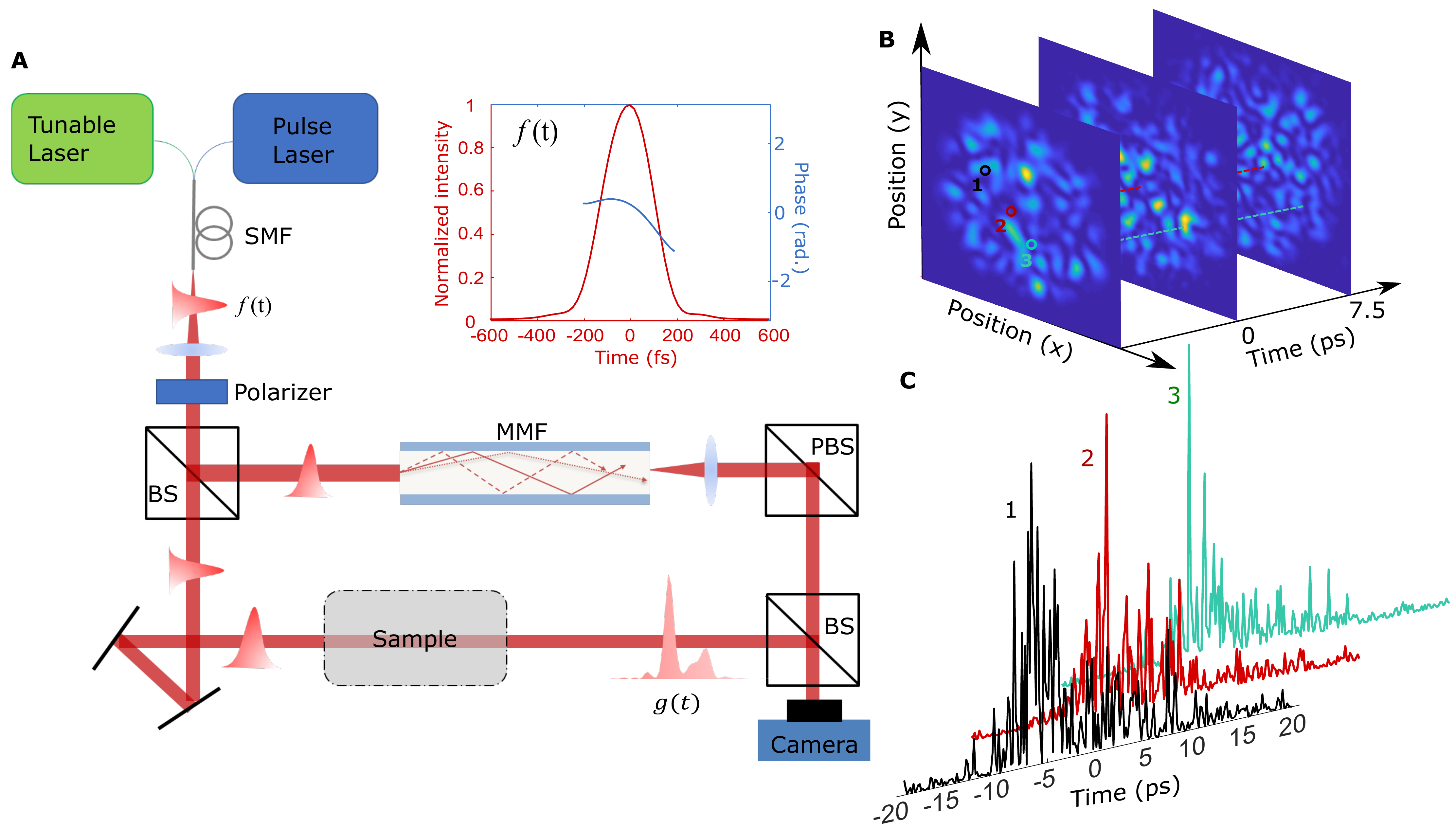}
        \caption{\textbf{Experimental setup and measured spatio-temporal speckles.} (\textbf{A}), Schematic of a Mach-Zehnder interferometric setup for full-field measurement. A CW laser with tunable frequency (Agilent 81940A) for MMF calibration and a pulsed laser (NKT, Onefive Origami) for temporal measurement are sequentially coupled into a single-mode fiber (SMF). The output is collimated by a lens, and one polarization is selected by a linear polarizer. The beam is split by a beam splitter into the fiber arm and the reference arm with equal path length. Light fields from the two arms recombine by a second beam splitter, and their interference pattern is recorded by a camera (Xenics Xeva 1.7-640). Inset: intensity (red, solid) and phase (blue, solid) of the reference pulse of launched into the MMF. BS: beam splitter. PBS: polarizing beam splitter. (\textbf{B}), Spatial field (amplitude) distribution of the laser pulse transmitted through the MMF at three arrival times -7.5, 0 and 7.5 ps. (\textbf{C}), Temporal field amplitudes at three spatial positions of the fiber output facet, marked by matching colors in \textbf{B}.} 
        \label{fig:figure2}
    \end{figure*}
    
    \section{Result}
    The proposed scheme is experimentally demonstrated in a Mach-Zehnder interferometric setup shown schematically in Fig.~\ref{fig:figure2}A. A 230-fs-long pulse from a mode-locked near-IR fiber laser (NKT, Onefive Origami) is split by a beam splitter into two paths, one is launched into the MMF for creation of the spatio-temporal speckle field $E(\mathbf{r},t)$, and the other is sent to probe a sample placed in the other arm of the interferometer. The transmitted or reflected field $g(t)$ from the sample is combined with $E(\mathbf{r},t)$ by a second beam splitter. Since they are mutually coherent, they will interfere, as long as $g(t)$ overlaps with $E(\mathbf{r},t)$ in time, which is ensured by matching the optical path lengths of the two arms of the interferometer. To increase the temporal length of $E(\mathbf{r},t)$, which determines the measurement range, we adjust the launch condition for the reference pulse into the MMF so that it excites many guided modes that propagate at different speeds. Due to modal dispersion, the transmitted pulse is broadened and distorted. The pulse shape varies spatially across the fiber facet. To have strong modal dispersion, we choose a step-index fiber (105 $\mu$m core, 0.22 NA, Thorlabs FG105LCA) of 1.8-meter length. 
    
    To calibrate the spatiotemporal speckle at the fiber output, we measure the transmitted field profile in the spectral domain with a tunable continuous-wave (CW) fiber laser. The laser wavelength $\lambda$ is scanned from $1520$ nm to $1570$ nm with a step of 0.2 nm. This range fully covers the spectrum of the reference pulse, which is centered at $\lambda = 1546$ nm and has a full width at half maximum (FWHM) $\Delta \lambda = 12$ nm, as reported in the Supplementary Materials, Section S1. To ensure the launch condition of the CW laser light into the MMF is identical to that of the pulsed laser, the outputs from both lasers are coupled into a single-mode fiber (SMF) switch. The CW light transmitted through the MMF is combined with that from the reference arm at a slight angle, producing spatial interference fringes. By applying a Hilbert filter in the Fourier domain of the recorded interference pattern, the amplitude and phase of the transmitted field at a single frequency $\omega$ are extracted. Scanning the frequency $\omega$ of the CW laser and repeating this off-axis holography measurement gives the frequency-resolved field transmission matrix $T(\mathbf{r}, \omega)$. At both the input and the output of the MMF, only one polarization is selected.

    \begin{figure*}[t]
        \includegraphics[width=2\columnwidth,keepaspectratio,clip]{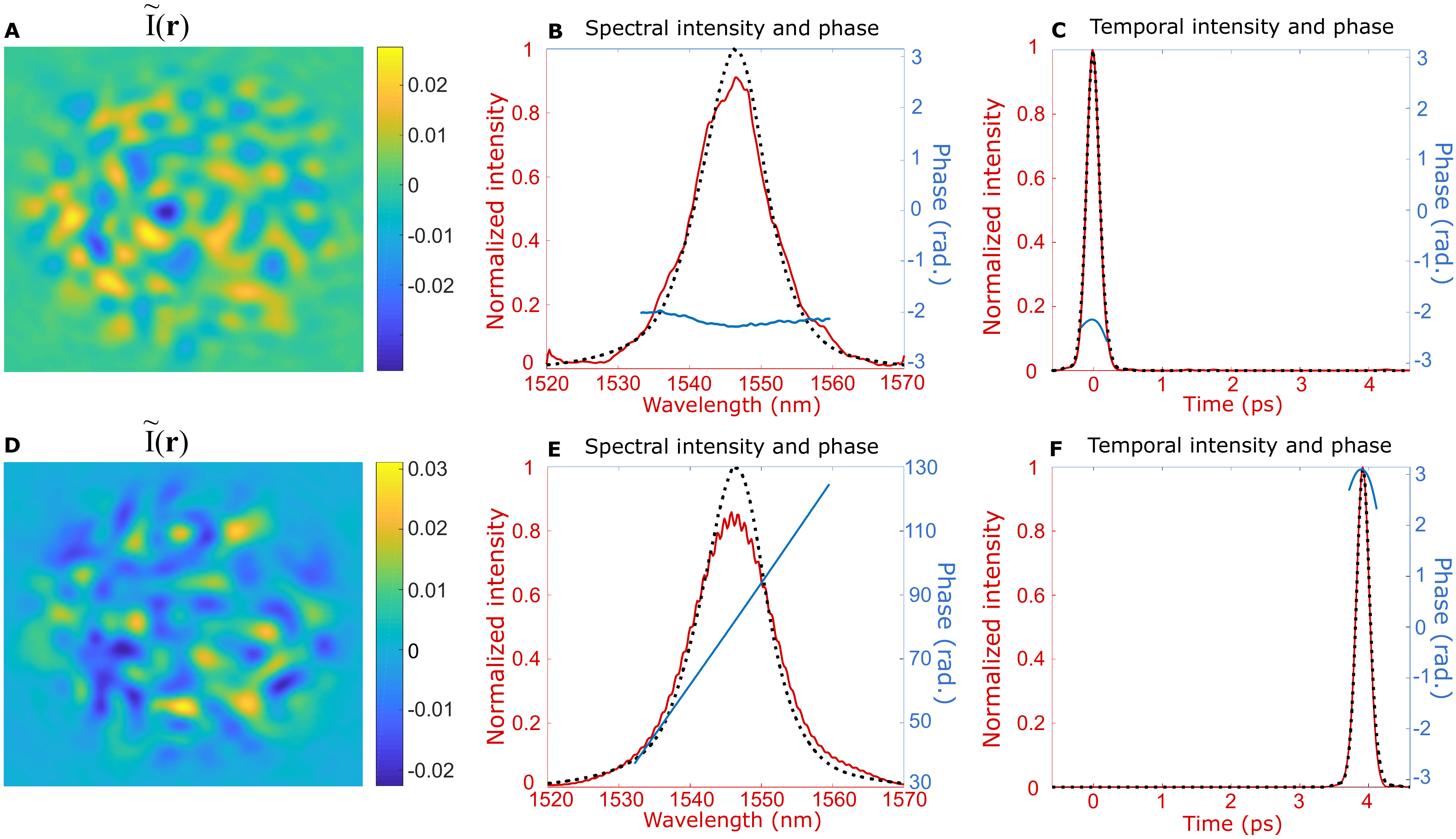}
        \caption{\textbf{Full-field measurement of single pulses with varying delay.} (\textbf{A, D}) 2D interference term $\tilde{I}(\mathbf{r})$ extracted from the off-axis hologram for a single pulse with arrival times $\tau$ = 0 and 4 ps. (\textbf{B, E}) Spectral intensity (red solid line, left axis) and spectral phase (blue solid line, right axis) of the signals retrieved from (\textbf{A, D}). The black dotted line is the spectral intensity of the signal measured by an optical spectrum analyzer. (\textbf{C, F}) Temporal intensity (red solid line, left axis) and temporal phase (blue solid line, right axis) of the signals, obtained by a Fourier transform of (\textbf{B, E}) respectively. Black dotted line is the temporal intensity of the signal obtained from autocorrelation and spectrum measurements.}
        \label{fig:figure3}
    \end{figure*}
    
    \begin{figure*}[t]
        \includegraphics[width=2\columnwidth,keepaspectratio,clip]{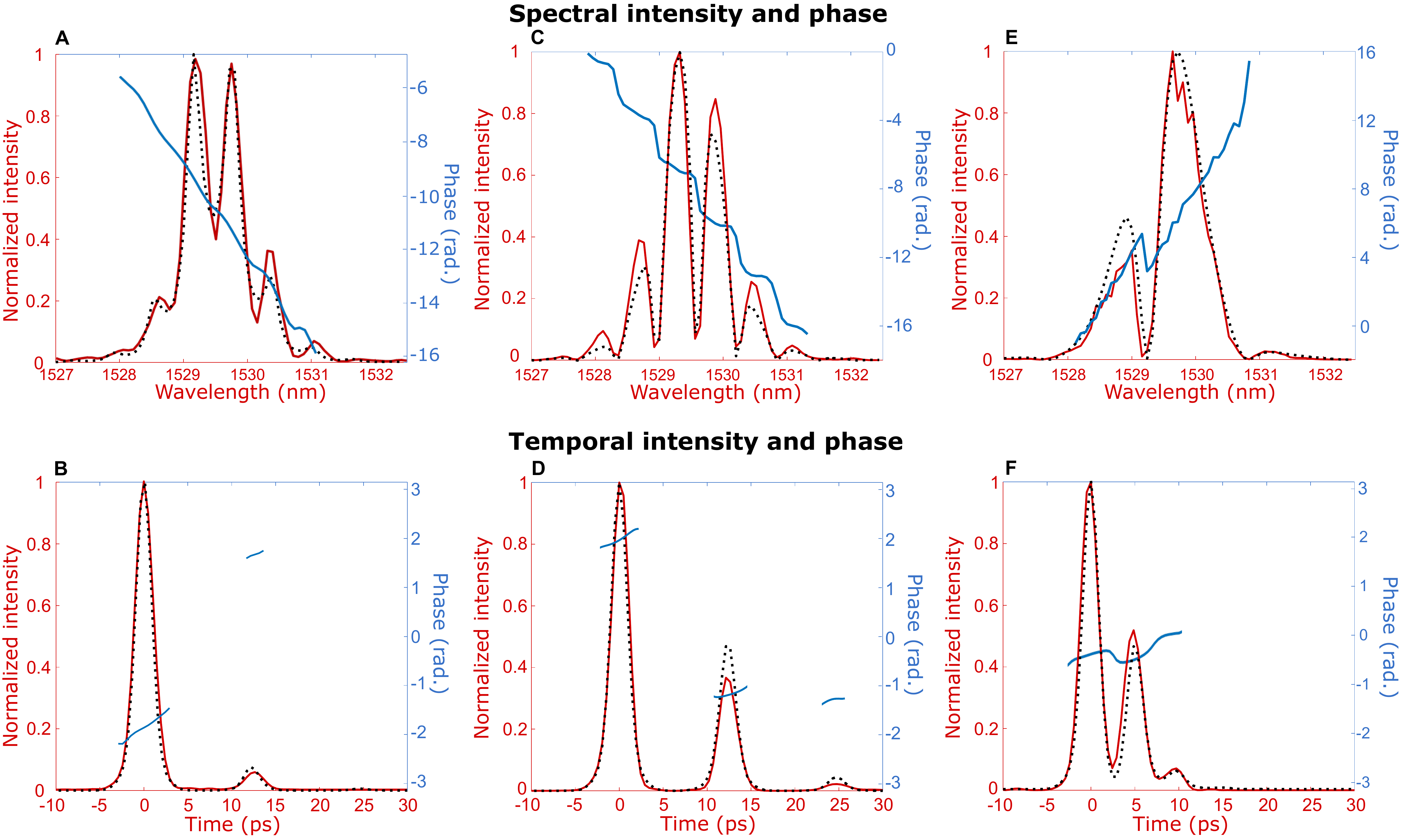}
        \caption{\textbf{Full-field measurement of multiple pulses.} First row: spectral intensity (left axis) and spectral phase (right axis). Second row: temporal intensity (left axis) and temporal phase (right axis). (\textbf{A, B}) Transmission of the reference pulse through a silicon wafer of thickness 535 $\mu$m with an incident angle of $1^{\circ}$. (\textbf{C, D}) Reflection of the reference pulse from the 535 $\mu$m-thick silicon wafer at the incident angle of $3^{\circ}$. (\textbf{E, F}) Reflection of the reference pulse from a 212 $\mu$m-thick silicon wafer at an incident angle of $2^{\circ}$. } 
        \label{fig:figure4}
    \end{figure*}
    
    After calibrating $T(\mathbf{r},\omega)$ of the MMF with the tunable CW laser, the input source is switched to a pulsed laser. The reference pulses are characterized by an autocorrelator and an optical spectrum analyzer (see details in the Supplementary Materials, Section S1). From these measurements, the spectral amplitude and phase of the reference field $f(\omega)$ are obtained. The Fourier transform $\mathcal{F}$ gives the temporal waveform $f(t) = \mathcal{F}[f(\omega)]$ of the reference pulse that is injected into the MMF. The temporal shape and phase of the reference pulse are shown in the inset of Fig.~\ref{fig:figure2}A. The transmitted field of the MMF is $E(\mathbf{r},\omega) = T(\mathbf{r},\omega)f(\omega)$ in frequency, and $E(\mathbf{r},t) = \mathcal{F}[E(\mathbf{r},\omega)]$ in time. As shown in Fig.~\ref{fig:figure2}B, the output speckle pattern changes rapidly in time. At each spatial location, the distinct temporal waveform is composed of multiple speckles, as plotted in Fig.~\ref{fig:figure2}C. 
    
    The complex yet deterministic spatiotemporal speckles generated by the MMF enable single-shot full-field measurement of the unknown signal by interfering $E(\mathbf{r},t)$ and $g(t)$. The time-integrated interference pattern is recored by off-axis holography, $I(\mathbf{r}) = \int |E(\mathbf{r},t)+g(t)|^2 dt$. The information of $g(t)$ is encoded in the interference term $\tilde{I}(\mathbf{r}) = \int dt [E(\mathbf{r},t)g^*(t) +  E(\mathbf{r},t)^*g(t)]$. $\tilde{I}(\mathbf{r})$ is extracted with the same Hilbert filter used in the calibration of $T(\mathbf{r},\omega)$. In the frequency domain, the interference term is $\tilde{I}(\mathbf{r}) = \int [T(\mathbf{r},\omega)f(\omega)g^*(\omega) + T^*(\mathbf{r},\omega)f^*(\omega)g(\omega)]d\omega$, where $g(\omega)$ is the Fourier transform of $g(t)$. This expression can be rewritten as a linear product of a matrix and a vector:
    \begin{equation}
    \tilde{I}(\mathbf{r}) =  \begin{bmatrix}
    T(\mathbf{r},\omega)f(\omega) &
    T^*(\mathbf{r},\omega)f^*(\omega)
    \end{bmatrix}
    \begin{bmatrix}
    g^*(\omega)\\
    g(\omega)
    \end{bmatrix}.
    \label{eq:eq1}
    \end{equation}
    With $T(\mathbf{r},\omega)$ and $f(\omega)$ known, $g(\omega)$ is retrieved from $\tilde{I}(\mathbf{r})$ by an iterative optimization algorithms. To account for temporal or spectral sparsity of $g(\omega)$, we deploy a compressive sensing algorithm FASTA \cite{Goldstein14} to solve the sparse least square optimization problem.
    
    To find the temporal resolution, we compute the temporal correlation function of the spatiotemporal speckle field, $C(\Delta t) \equiv \langle E^*(\mathbf{r},t) E(\mathbf{r},t+ \Delta t) \rangle$, where $\langle ... \rangle$ denotes averaging over $\mathbf{r} $ and $t$. The FWHM of $C(\Delta t)$ gives the average temporal speckle size $\delta t = 230 $ fs, which determines the temporal resolution.  
    
    The temporal range of measurement $\Delta t$ is equal to the temporal length of $E(\mathbf{r},t)$, which is inversely proportional to the width of the spectral correlation function $C(\Delta \omega) \equiv \langle E^*(\mathbf{r},\omega) E(\mathbf{r},\omega + \Delta \omega) \rangle$. From the width of $C(\Delta \omega)$, we estimate $\Delta t$ to be about 35 ps. The time bandwidth product (TBP), defined by the ratio of the temporal range to the temporal resolution, is $\Delta t / \delta t = 152 $.

    We first test our method by measuring single pulses propagating through the reference arm (without a sample) of the Mach-Zehnder interferometer with different delay times. By changing the length of the reference arm with a delay line, we vary the arrival time $\tau$ of the pulse. The zero delay time $\tau = 0$ is set by the arrival time of the pulse when the length of the reference arm is matched to that of the fiber arm. Figure~\ref{fig:figure3} shows the measurement results for two delay times $\tau= 0$ (top row) and $\tau = 4$ ps (bottom row). The left column shows the interference term  $\tilde{I}(\mathbf{r})$ extracted from the experimentally measured hologram in these two cases. Although the pulse shape remains the same, the spatial interference pattern is very different. It is because pulses with varying delays interfere with different parts of the spatiotemporal speckles from the MMF. The retrieved intensity and phase of the pulses in wavelength and time domains are plotted in the second and third columns. The recovered spectral intensity is consistent with the measurement using an optical spectral analyzer. While the recovered spectral phase is flat for $\tau= 0$, it changes linearly for $\tau= 4$ ps. These results are expected, as the slope of the spectral phase corresponds to the delay time. In the time domain, the arrival times of the recovered pulses agree with the values set by the delay line, and the temporal pulse shape is consistent with the autocorrelation trace (see Section S1 in the Supplementary Materials).

    We next measure double pulses. Unstable double-pulsing is a common phenomenon for lasers that are over-pumped, but it is difficult to detect with repetitive measurement techniques that rely on stable pulse trains. Our method can measure double pulses in a single shot. To produce double pulses, we first create 2.2-ps-long pulses by spectral filtering the output from a mode-locked fiber laser (Calmar Mendocino) (see the Supplementary Materials, Section S1). Then we insert a double-side-polished silicon wafer to the reference arm of the March-Zehnder interferometer as the sample. An incident pulse is reflected back and forth between the two surfaces of the wafer, creating multiple pulses in transmission. The distance between the pulses is determined by the wafer thickness and the incident angle. Fig.~\ref{fig:figure4}A shows the recovered spectral intensity, which exhibits a rapid oscillation. To verify the result, we also simulate the transmission spectrum of the silicon wafer using the transfer matrix method. The wafer thickness is 535 $\mu$m and the incident angle of the probe pulse is 1 degree (see the Supplementary Materials, Section S2).  The simulated spectrum (black dotted line) agrees well with the recovered spectrum (red solid line). The recovered spectral phase, unwrapped and plotted by the blue solid line, features descending jumps at the frequencies of local minima for the spectral intensity. These phase jumps, together with the amplitude oscillations, are results of spectral interference of the double pulse, which is reconstructed from the Fourier transform of the recovered spectral field in Fig.~\ref{fig:figure4}B. The first pulse originates from direct transmission of the probe pulse through the wafer and the second pulse from two reflections within the wafer. They are spaced by 12.5 ps, which is consistent with the 3.75-mm-long one-round-trip optical path length. Because of the relatively low reflectivity of the silicon-air interface, the intensity ratio of the first pulse to the second pulse is 17.6. Although the second pulse is rather weak, it can still be recovered by our scheme, and the temporal shape agrees well with the simulation result. The temporal phases of the two pulses vary linearly, reflecting the absence of frequency chirp within each pulse.
    
    Finally, we measure more complex pulses that are created by reflection from a silicon wafer. The interferometric setup is slightly modified to measure the pulses reflected by the sample in the reference arm (see the Supplementary Materials, Section S3). The recovered spectral intensity in Fig.~\ref{fig:figure4}C features interference fringes of higher contrast, and the spectral phase exhibits larger jumps than those in Fig.~\ref{fig:figure4}A.  The Fourier transform of the recovered spectral field reveals three pulses in the time domain, as plotted in Figure~\ref{fig:figure4}. The first pulse results from the direct reflection of the incident pulse by the front surface of the wafer, and the second pulse from direct reflection by the back surface. The intensity ratio of the second pulse to the first pulse in reflection is higher than that in transmission, leading to more pronounced interference fringes in the spectral domain. Nevertheless, the frequency spacing of the fringes, which is determined by the delay time between the two pulses, remains the same. Even the third pulse, generated by three reflections in the wafer, is still visible and recovered in our measurement. The recovered spectral and temporal intensities are in good agreement with the simulation results. The temporal phase within each of the three pulses increases linearly in time with the same slope, indicating each pulse has the same frequency and no chirp. 
    
    To reduce the pulse spacing, we switch to a second wafer of a smaller thickness (212 $\mu$m). The reconstructed pulses in reflection are shown in Fig.~\ref{fig:figure4}F. The delay time between adjacent pulses is reduced to 5 ps, corresponding to one round-trip in the silicon wafer. However, the relative intensities of the three pulses are not changed, as they are determined by the reflectivity of the silicon-air interface. The shorter delay time corresponds to the larger spacing of spectral fringes, leading to a smaller number of fringes within the same frequency range in Fig.~\ref{fig:figure4}E. The recovered pulses have a slight overlap in time, again in good agreement with the simulation result.  
    
    \section{Discussion}
    In summary, we demonstrate a novel MMF-based scheme for the single-shot full-field measurement of complex pulses. We obtain a temporal resolution of $230$ fs, a temporal range of $\sim$35 ps and a TBP of 152. The temporal resolution can be further enhanced by increasing the spectral bandwidth of the reference pulse. The temporal range of the single-shot measurement, which is governed by the spectral correlation width of the MMF, may be tuned independently of the temporal resolution. For example, using a 100-meter-long MMF will increase the temporal range to nanosecond \cite{Redding14}.  The TBP is limited by the number of guided modes in the MMF, which may well exceed 1000 for a fiber with a large core and a high numerical aperture. Using a bundle of MMFs will further increase the TBP \cite{Liew16}.  
    
    Taking full advantage of the complex spatio-temporal speckles created by the reference pulse through an MMF, our scheme eliminates the mechanical scanning of the time delay between the signal and the reference. Furthermore, our method overcomes the limitation of spectral resolutions in spectral interferometry \cite{Froehly73, Lepetit95,Dorrer99,Dorrer00}. Comparing to other single-shot methods based on nonlinear processes such as time lens \cite{Iaconis98,shea01,Li17,Ryczkowski18,Tikan18}, our scheme is based on linear interferometry, which possesses a much higher sensitivity. With the knowledge of the reference pulse, as required by all linear interferometric methods \cite{Monmayrant10}, it can measure non-reproducible and non-periodic ultra-weak signals. The reference pulse is not necessarily transform-limited, as long as its temporal/spectral amplitude and phase are known. Even without knowledge of the reference, the relative phase and amplitude change imposed by the sample can still be recovered (see Section S4 in the Supplementary Materials). The simplicity and high sensitivity of our method illustrate the potential of MMFs as versatile and multi-functional sensors. 

\section{Supplementary Materials}
Supplementary materials for this article are available at xxxx.

Section S1. Calibrations of reference pulses.

Section S2. Characterizations of silicon wafer samples.

Section S3. Setup of the reflection measurement.

Section S4. Measuring sample properties without a known reference pulse.

Fig. S1. Calibration of the NKT laser.

Fig. S2. Calibration of the Calmar laser.

Fig. S3. Transmission and reflection spectra of two silicon wafers.

Fig. S4. Schematics of the reflection measurement setup.

\section{Acknowledgment}
The authors acknowledge Chia Wei Hsu, Yaniv Eliezer, Logan Wright for fruitful discussions. 

\textbf{Funding:} This work is supported by US National Science Foundation under the Grant Nos. ECCS-1509361 and ECCS-1809099.

\textbf{Author contributions:} W. X conducted the experiment, simulation, and computational algorithms. S. G. contributed to the development of algorithms. H. Y. prepared the wafer samples and contributed to the measurement. H. C. initiated and supervised the project. W. X. and H. C wrote the manuscript with feedback from S. G. and H. Y.  

\textbf{Competing interests:} The authors declare that they have no competing interests. 

\textbf{Data availability:} All data necessary to evaluate the conclusions in the paper are present in the paper and/or the Supplementary Materials. Additional data related to this paper may be requested from the authors.

\clearpage
\clearpage


\section*{Supplementary material}

# Multimode fiber based single-shot full-field measurement of optical pulses: Supplementary Materials

Wen Xiong, Shai Gertler, Hasan Yılmaz, and Hui Cao*

*Department of Applied Physics, Yale University, New Haven, Connecticut 06520, USA*

(Dated: July 21, 2019)

## I. PULSE CALIBRATION

We have used two mode-locked fiber lasers to generate reference pulses. The first one is NKT Onefive Origami. When the diode current is set to 750 mA, the emission spectrum is centered at wavelength $\lambda = 1546$ nm and has a full width at half maximum (FWHM) of 12 nm. The spectrum of the pulse is measured by an optical spectrum analyzer (OSA, YOKOGAWA AQ6370D), and plotted in Fig. S1**a**. To obtain the temporal pulse shape, we measure the temporal autocorrelation trace of the pulse with an autocorrelator (Femtochrome FR-103XL), see Fig. S1**b**. We fit the spectral phase of the pulse (blue dashed line in Fig. S1**a**) so that the corresponding autocorrelation trace (black dashed line in Fig. S1**b**) matches well with the measured one. The FWHM of the pulse is 230 fs, as shown in the inset of Fig. 2a of the main text.

The second pulsed laser is Calmar Mendocino. By spectral filtering of the laser output, we obtain longer pulses. We use a fiber-based tunable filter and first set the center wavelength to 1532 nm and the FWHM to 10 nm. The measured spectrum of the filtered pulse is plotted in Fig. S2**a**, and the autocorrelation trace is shown in Fig. S2**b**. Again we fit the spectral phase of the filtered pulse (blue dashed line in Fig. S2**a**) so that the autocorrelation traces agree well. Then we further reduce the FWHM of the spectral filter to 1 nm. The corresponding spectrum measured by the OSA is shown in Fig. S2**c**. The spectral phase within this narrow range is nearly constant so that the filtered pulse is almost transform-limited and has a length of 2.2 ps.

## II. SAMPLE CHARACTERIZATION

The samples used in the temporal measurements described in the main text are two silicon wafers of thicknesses approximately 500 $\mu$m and 200 $\mu$m. To find the precise thickness of the samples and the incident angle of the probe light, we measure the transmission spectrum of the silicon wafer with the experimental setup shown in Fig. 2**a** in the main text. The wafer is placed in the reference arm and the fiber arm is blocked. By scanning the frequency of a CW laser, we measure the spatially integrated intensity of transmitted light with the camera as a function of wavelength. Then we simulate the transmission spectrum of a silicon wafer with the transfer matrix method. By matching the simulated spectrum (black dashed line in Fig. S3**a**) to the measured one (red solid line), we find that the wafer thickness is 535 $\mu$m and the incident angle is 1°.

## III. REFLECTION MEASUREMENT

To measure the reflection from the sample, the experimental setup is slightly modified, as shown schematically in Fig. S4. The reference arm is divided into two by a beam splitter (BS), one is the reflection from a mirror (M3), the other from the sample. When calibrating the field transmission matrix of the multimode fiber, a beam blocker is placed in front of the sample, and the reflection from the mirror M3 serves as the reference for off-axis holography. To measure the reflection from the sample, the beam blocker is placed in front of the mirror M3.

We use the tunable CW laser to measure the reflection spectrum of the silicon wafers, as plotted by red solid lines in Fig. S4**b-c**. By matching them to the transfer matrix simulation results (black dashed line), we find that the two wafers are 535 $\mu$m and 212 $\mu$m thick, and the angles of incidence are 3° and 2°, respectively.

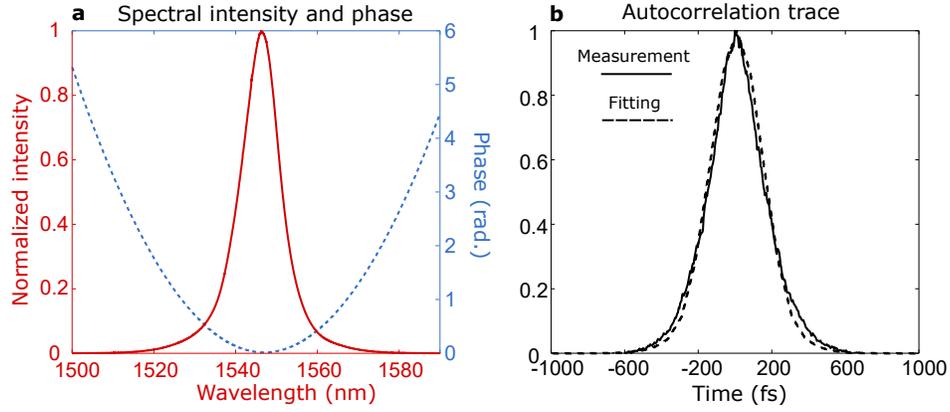

FIG. S1: Calibration of the femtosecond reference pulse $f(t)$. **a.** Red solid line is the measured spectral intensity of the laser pulse from NKT Onefive Origami, and the blue dashed line is the spectral phase obtained from fitting the measured temporal autocorrelation trace of the pulse in **b**. **b.** Measured temporal autocorrelation trace of the pulse (black solid line) and simulated trace (black dashed line) from the measured spectral intensity and fitted spectral phase in **a**.

## IV. CHARACTERIZING SAMPLE RESPONSE

Even if the reference pulse is unknown, our method can extract spectral/temporal response of the sample. Eq. (1) in the main text can be rewritten as

$$\tilde{I}(\mathbf{r}) = \begin{bmatrix} T(\mathbf{r}, \omega) & T^*(\mathbf{r}, \omega) \end{bmatrix} \begin{bmatrix} f(\omega)g^*(\omega) \\ f^*(\omega)g(\omega) \end{bmatrix}. \tag{S1}$$

From the measured interference pattern $\tilde{I}(\mathbf{r})$, $f^*(\omega)g(\omega)$ is recovered. Then we remove the sample from the reference arm of the interferometer and repeat the measurement to get $f^*(\omega)f(\omega)$. The ratio $f^*(\omega)g(\omega)/f^*(\omega)f(\omega) = g(\omega)/f(\omega) = h(\omega)$ gives the spectral response of the sample, and its Fourier transform gives the temporal response $h(t)$.

---

$^*$ Electronic address: hui.cao@yale.edu

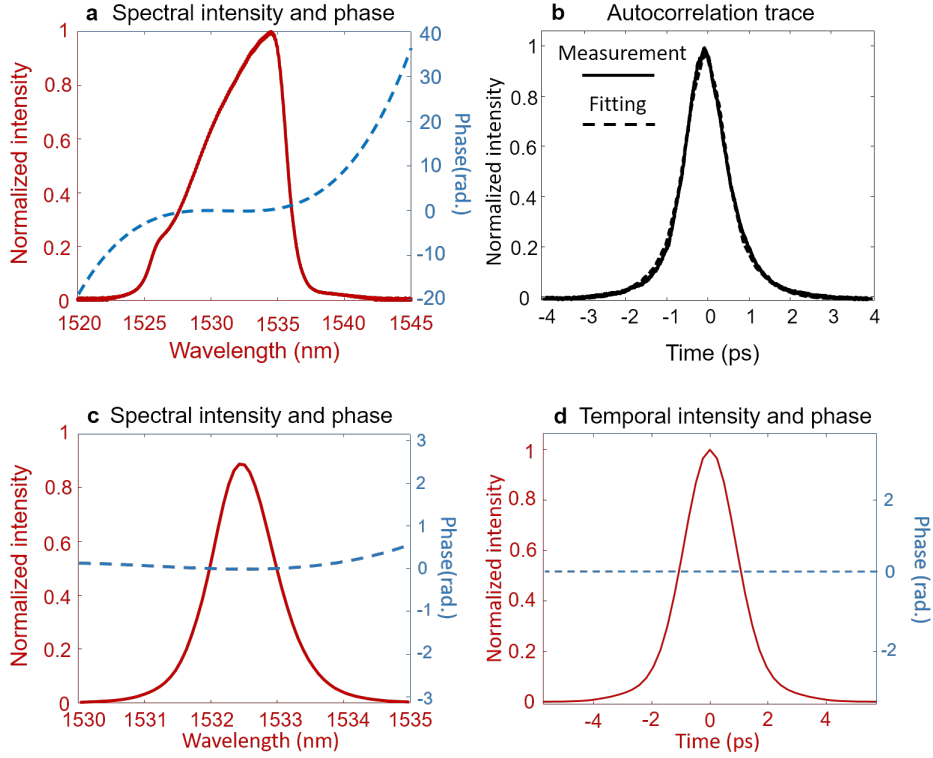

FIG. S2: Calibration of the picosecond reference pulse. **a.** Red solid line is the measured spectral intensity of the laser pulse after being spectrally filtered to 10 nm FWHM at 1532 nm, blue dashed line is the spectral phase obtained from fitting the temporal autocorrelation trace in **b**. **b.** Measured temporal autocorrelation trace (black solid line) and simulated trace (black dashed line) from the measured spectral intensity and fitted spectral phase in **a**. **c.** Measured spectral intensity (red solid line) of the filtered pulse with the FWHM reduced to 1 nm. The spectral phase (blue dashed line), obtained from **b**, is nearly flat. **d.** Temporal intensity (red solid line) and phase (blue dashed line) of the filtered pulse used as the reference.

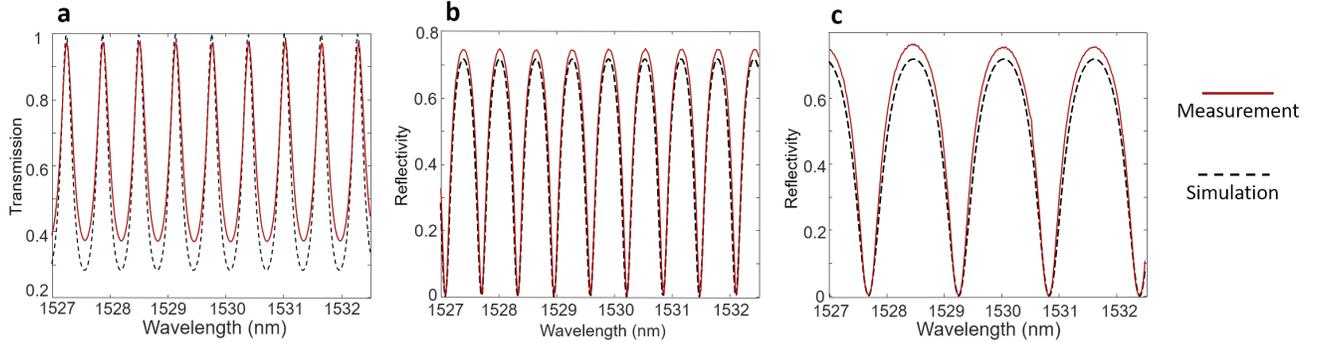

FIG. S3: Transmission and reflection spectra of the two silicon wafers. Red solid lines: experimental data. Black dashed lines: numerical results from transfer matrix simulation. **a.** The transmission spectrum of the 535 $\mu$m silicon wafer with an incident angle of 1 degree. **b.** The reflection spectrum of the 535 $\mu$m wafer with an incident angle of 3 degrees. **c.** The reflection spectrum of the 212 $\mu$m wafer with an incident angle of 2 degrees.

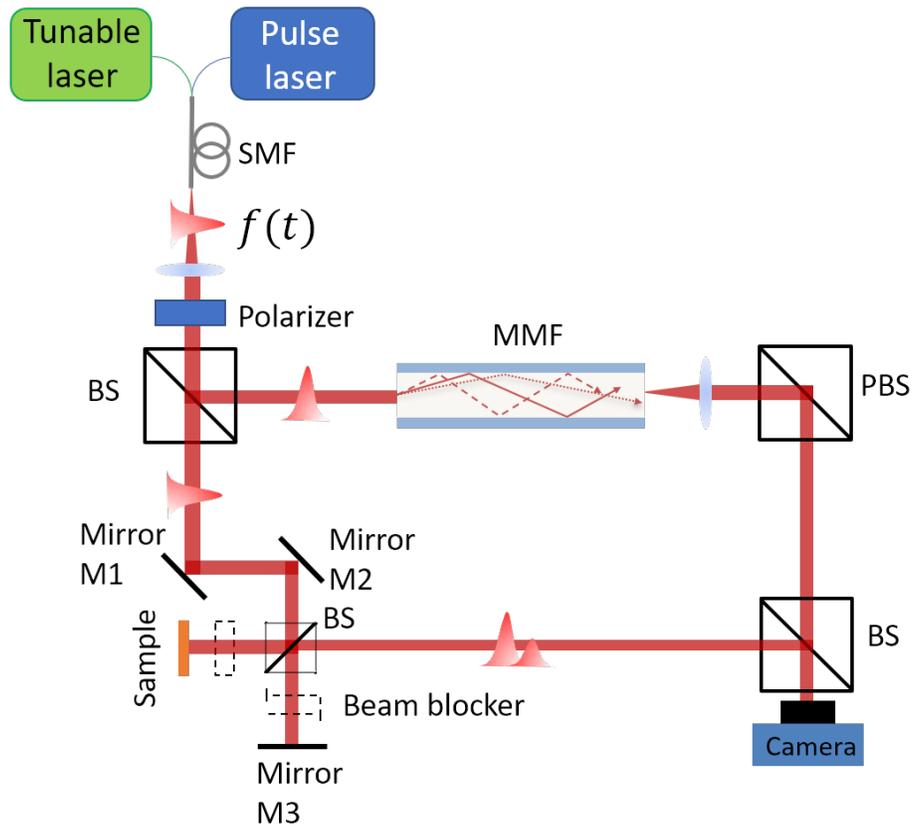

FIG. S4: Experimental setup for measuring the reflected pulses from the sample. The reference arm of the March-Zehnder interferometer is slightly modified from the one shown in Fig. 2**a** in the main text. The reference arm is divided into two by a beam splitter (BS). The beam blocker blocks the reflection from the sample during the calibration of the fiber transmission matrix, and it blocks the reflection from the mirror M3 in the measurement of reflection from the sample. The positions of the mirror M3 and the sample are adjusted to match the optical path-length to that of the fiber arm.


\end{document}